\documentclass[%
reprint,
groupedaddress,
amsmath,amssymb,
aps,
prb,
superscriptaddress]{revtex4-1}
\usepackage{graphicx}% Include figure files
\usepackage{dcolumn}
\usepackage{bm}
\usepackage{physics}
\usepackage{color}

\bibpunct{[}{]}{,}{n}{}{}

\RequirePackage[colorlinks=true,linkcolor=blue,urlcolor=blue,hyperfootnotes=false,citecolor=blue]{hyperref}

\begin{document}
\title{Engineering Entangled Coherent States of Magnons and Phonons via a Transmon Qubit}

\author{Marios Kounalakis}
\email{marios.kounalakis@gmail.com}

\affiliation{Kavli Institute of Nanoscience, Delft University of Technology, 2628 CJ Delft, The Netherlands}
\affiliation{Institute for Theoretical Solid State Physics, RWTH Aachen University, 52074 Aachen, Germany}
\author{Silvia Viola Kusminskiy}
\affiliation{Institute for Theoretical Solid State Physics, RWTH Aachen University, 52074 Aachen, Germany}
\author{Yaroslav M. Blanter}
\affiliation{Kavli Institute of Nanoscience, Delft University of Technology, 2628 CJ Delft, The Netherlands}
\date{\today}

\begin{abstract}
We propose a scheme for generating and controlling entangled coherent states (ECS) of magnons, i.e. the quanta of the collective spin excitations in magnetic systems, or phonons in mechanical resonators.
The proposed hybrid circuit architecture comprises a superconducting transmon qubit coupled to a pair of magnonic Yttrium Iron Garnet (YIG) spherical resonators or mechanical beam resonators via flux-mediated interactions.
Specifically, the coupling results from the magnetic/mechanical quantum fluctuations modulating the qubit inductor, formed by a superconducting quantum interference device (SQUID).
We show that the resulting radiation-pressure interaction of the qubit with each mode, can be employed to generate maximally-entangled states of magnons or phonons.
In addition, we numerically demonstrate a protocol for the preparation of magnonic and mechanical Bell states with high fidelity including realistic dissipation mechanisms.
Furthermore, we have devised a scheme for reading out the prepared states using standard qubit control and resonator field displacements.
Our work demonstrates an alternative platform for quantum information using ECS in hybrid magnonic and mechanical quantum networks.
\end{abstract}
\maketitle
% \twocolumngrid

\section{Introduction}
The development of quantum technologies aims towards disruptive practical applications in several fields such as computing, communication and sensing by exploiting the effects of quantum mechanics~\cite{macfarlane2003quantum,acin2018quantum}.
The success of this venture largely relies on the evolution of hybrid quantum systems that incorporate the advantages of different physical platforms in a constructive way~\cite{kurizki2015quantum,clerk2020hybrid}.
For example, circuit quantum electrodynamics (QED), where light-matter interactions in superconducting circuits are used to manipulate quantum information, is one of the leading platforms in quantum computing, combining strong nonlinearities with advanced quantum control and readout as well as high coherence times relative to qubit operations~\cite{devoret2013superconducting,kjaergaard2020superconducting}.
However, superconducting circuits do not directly couple to optical photons, hindering their integration with optical networks~\cite{kurizki2015quantum}. 
In this direction, the development of hybrid circuit QED platforms based on mechanical and magnetic systems is an essential requirement towards networked quantum computation~\cite{clerk2020hybrid}.
In addition, the evolution of high-quality mechanical systems operating in the quantum regime provides unique opportunities not only in transduction but also in building quantum memories and sensors~\cite{aspelmeyer2014cavity,kurizki2015quantum,clerk2020hybrid}.
Moreover, hybrid quantum systems based on magnons, i.e., the quanta of the collective spin excitations in magnetic materials, offer distinctive advantages, such as unidirectional propagation and chiral coupling to phonons and photons~\cite{yu2020magnon,bertelli2020magnetic}, making them prime candidates for technological applications in quantum information sciences~\cite{lachance2019hybrid,yuan2022quantum}.

The ability to generate entanglement is at the heart of most protocols in quantum information.
For macroscopic mechanical and magnonic resonators, which carry bosonic degrees of freedom and typically operate in the linear regime, the special class of entangled coherent states (ECS)~\cite{yurke1986generating,sanders2012review} is of particular interest.
Such states exhibit continuous-variable entanglement between different bosonic modes and provide a valuable resource for quantum teleportation~\cite{enk2001entangled,wang2001quantum}, quantum computation~\cite{cochrane1999macroscopically,oliveira2000quantum,jeong2002efficient} and communication~\cite{loock2008hybrid,allati2011communication}.
In addition, ECS are useful for fundamental studies of quantum mechanics with applications in quantum metrology~\cite{munro2002weak,joo2011quantum} and tests of collapse models~\cite{diosi2011gravity,kafri2014classical}.

% This is typically achieved using yttrium iron garnet (YIG) spherical resonators, which exhibit higher magnon lifetimes~\cite{tabuchi2014hybridizing}.

% The addition of superconducting qubits in such hybrid schemes unlocks a wide range of possibilities such as arbitrary quantum control and detection of magnons and phonons~\cite{tabuchi2015coherent,reed2017faithful,chu2018creation,sletten2019resolving,lachance2020entanglement,xu2022quantum,kounalakis2019synthesizing,sharma2022protocol,kounalakis2022analog}.

Macroscopic entanglement between mechanical modes has recently been achieved on aluminum drum resonators~\cite{ockeloen2018stabilized,kotler2021direct} and micromechanical photonic/phononic crystal cavities~\cite{riedinger2018remote,wollack2022quantum}, however, an experimental demonstration of entanglement between metallic nanobeams such as the ones studied in Refs.~\cite{rodrigues2019coupling,schmidt2020sideband,bera2021large} is currently lacking.
Furthermore, while entanglement between atomic ensembles has been experimentally realised in an optical setup~\cite{simon2007single}, entangling magnons in two distant magnets still remains a challenge.
Recent theoretical proposals have investigated the possibility of entangling magnons in two Yttrium Iron Garnet (YIG) spheres interacting via photons in a microwave cavity.
More specifically, in Ref.~\cite{zhang2019quantum} the emerging Kerr nonlinearity in strongly driven magnons is used, relying on driving the magnon modes far from equilibrium in order to create entanglement.
In Ref.~\cite{li2019entangling} the nonlinearity stemming from the parametric magnetorestrictive interaction is employed to create magnon-magnon entanglement, although requiring a much larger magnetorestrictive coupling strength than experimentally attainable~\cite{zhang2016cavity}.
Alternatively, in Refs.~\cite{nair2020deterministic,yu2020macroscopic} it is shown that two YIG spheres can be entangled by driving the magnon-cavity system with strong squeezing fields.
However, while the above schemes show promise for creating magnon-magnon entanglement in distant YIG spheres, the absence of a highly controllable nonlinear element, such as a qubit, hinders the generation and control of more complex states and ECS, in particular.
% despite the potential for creating magnon-magnon entanglement in all these schemes, the possibility of generating and controlling ECS is hindered by the absence of a highly controllable nonlinear element such as a qubit.

Here we propose a scheme for generating ECS of magnons/phonons in a hybrid circuit QED architecture comprising a superconducting transmon qubit and two magnonic/mechanical modes.
Concerning magnonic systems, without loss of generality, we consider two YIG sphere modes in a hybrid qubit-magnon setup similar to Ref.~\cite{kounalakis2022analog}, where the qubit-magnon coupling is mediated via a superconducting quantum interference device (SQUID).
We showcase a protocol for generating maximally-entangled states such as \emph{Bell} and \emph{NOON} states with high fidelity, by exploiting the parametric nature of the qubit-magnon radiation-pressure interaction and the transmon quantum control toolbox.
Furthermore, we analyze a readout scheme for verifying the entanglement in the system based on qubit measurements and displacements of the magnon field.
Contrary to previous proposals for generating magnon-magnon entanglement, there is no need for placing the YIG spheres inside a cavity, therefore, increasing scalability and modularity.
Furthermore, we numerically demonstrate the validity of our proposal for entangling SQUID-embedded mechanical beam resonators~\cite{rodrigues2019coupling,schmidt2020sideband,bera2021large,kounalakis2020flux,khosla2018displacemon}, thereby extending the possibilities for quantum control using mechanical ECS.

\section{Hybrid system description}
\begin{figure}[t]
  \begin{center}
  \includegraphics[width=\linewidth]{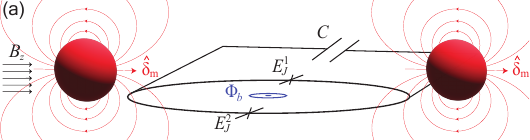}
   \includegraphics[width=\linewidth]{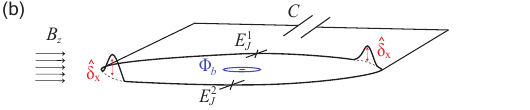}
 \end{center}
  \caption{
  {Proposed hybrid circuit architecture.
  A flux-tunable transmon qubit, formed by a C-shunted SQUID loop, is coupled to (a) two nearby YIG spheres or (b) two SQUID-embedded mechanical beams.
  The magnetization of both spheres in (a) is oriented by an in-plane field $B_z$.
  The magnetic quantum fluctuations $\hat{\delta}_m$ modulate the SQUID flux as well as the transmon inductive energy, thereby giving rise to a qubit-magnon coupling.
  In (b) the coupling stems from the mechanical quantum fluctuations $\hat{\delta}_x$ inducing a modulating flux in the SQUID in the presence of the in-plane field $B_z$.
  An additional flux bias $\Phi_b$ can be externally applied to tune the qubit frequency and modulate the coupling.}
  }
  \label{fig:scheme}
\end{figure}

% Our circuit architecture enables the coupling of several bosonic modes in magnetic and mechanical systems.
The fundamental element in the proposed circuit architecture is a dc SQUID, i.e., a superconducting loop interrupted by two Josephson junctions, as schematically depicted in Fig.~\ref{fig:scheme}.
When shunted by a capacitance $C$, with charging energy $E_C=2e^2/C$, this nonlinear inductor can realize a flux-tunable transmon qubit described by the Hamiltonian,
\begin{equation}
\hat{H}_T=4E_{C}\hat{N}^{2}-E_J\cos{\hat{\delta}},
\label{eq:CircuitHam}
\end{equation}
where $\hat{N}, \hat{\delta}$ are conjugate operators describing the tunneling Cooper-pairs and the superconducting phase across the SQUID, respectively~\cite{koch2007charge,vool2017introduction}.
In the case where the two junctions are the same (symmetric SQUID), an external flux bias $\Phi_b$ tunes the Josephson energy $E_J=E_J^\mathrm{max}\left\vert\cos{\phi_b}\right\vert$, where $\phi_b\dot{=}\pi\Phi_b/\Phi_0$ and $\Phi_0$ is the flux quantum.
%$E_J^\mathrm{max}$ is the sum of the Josephson energies of the two junctions.

For magnetic systems, without loss of generality we focus our description on micro-sized YIG spheres similar to Refs.~\cite{kounalakis2022analog,yuan2023magnon}.
Upon application of an in-plane magnetic field $B_z$, a YIG sphere acquires a magnetization $M_s$ and its excitations can be approximated as a set of independent quantum harmonic oscillators with Hamiltonian $\hat{H}_M=\hbar\sum_m\omega_{m}\hat{a}_m^{\dagger}\hat{a}_m$, where $a_m^{(\dagger)}$ are bosonic operators describing the annihilation (creation) of single magnons~\cite{tabuchi2016quantum,rameshti2021cavity}.
Note that this description is valid in the limit $\langle\hat{m}^{\dagger}\hat{m}\rangle\ll N_S$, where $N_S$ is the total number of spins in the sphere~\cite{tabuchi2016quantum,rameshti2021cavity}.
The fundamental excitation, or \emph{Kittel} mode, is a uniformly polarized state of all the spins acting as a single ``macrospin'' precessing around $z$, with ferromagnetic resonance (FMR) frequency $\omega_{0}=\gamma_{0}(B_{z}+B_\mathrm{ani})$, where $B_\mathrm{ani}$ is the anisotropy field~\cite{stancil2009spin}.
Higher mode frequencies are given by $\omega_{m}=\omega_{0}+\gamma_{0}M_s\frac{l-1}{3(2l+1)}$ depending on the magnon angular momentum quantum number $l$~\cite{sharma2019cavity}.

The mechanical systems of interest in this work consist of SQUID-embedded aluminum beams~\cite{rodrigues2019coupling,kounalakis2019synthesizing,kounalakis2020flux,schmidt2020sideband}.
Such mechanical beams are realised by suspending part of the SQUID loop such that it can freely oscillate out of plane~\cite{rodrigues2019coupling,schmidt2020sideband}.
Similar to the YIG sphere, its excitations can also be described by a set of independent quantum harmonic oscillators, with Hamiltonian $\hat{H}_{X}=\hbar\sum_x\omega_x\hat{a}_x^{\dagger}\hat{a}_x$, where $a_x^{(\dagger)}$ are bosonic operators that annihilate (create) a phonon.
The fundamental mode, which is the one considered in this work, oscillates with frequency $\omega_0=\hbar/(2mx_\mathrm{zpf}^2)$, where $m$ is the beam mass and $x_\mathrm{zpf}$ the magnitude of its zero-point motion~\cite{rodrigues2019coupling}.

Upon application of an in-plane magnetic field $B_z$, the quantum fluctuations in the out-of-plane displacement of the beam ${\hat{\delta}_x=x_\mathrm{zpf}(\hat{a}_x+\hat{a}_x^{\dagger})}$ induce a flux $\Phi(\hat{\delta}_x)=\beta_0B_zl\hat{\delta}_x$ through the loop, where $l$ is the beam length and $\beta_0$ is a geometric factor that depends on the mode shape~\cite{rodrigues2019coupling}.
Similarly, quantum fluctuations of the magnetic moment in the magnetized YIG sphere, $\hat{\delta}_m=\mu_\mathrm{zpf}(\hat{a}_m+\hat{a}_m^{\dagger})$, result in an additional flux $\Phi(\hat{\delta}_m)$ through the SQUID loop.
Let us assume that the sphere is placed at an in-plane and out-of-plane distance $d$ from the closest point in the loop.
Then in the far-field limit $\Phi(\hat{\delta}_m)=\mu_0\hat{\delta}_m/(4\sqrt{2}\pi d)$~\cite{kounalakis2022analog}.

The additional flux from each source of quantum fluctuation, $\Phi(\hat{\delta}_j)$, modulates the SQUID flux and consequently its Josephson energy,
\begin{equation}
E_J^\prime(\phi_b,\hat{\delta}_j)\simeq E_J\left(1-\tan{\phi_b}\sum_j\phi(\hat{\delta}_j)\right),
\label{eq:Ejexpanded}
\end{equation}
where we assume $\phi(\hat{\delta}_j)\dot{=}\pi\Phi(\hat{\delta}_j)/\Phi_0\ll1$ and a symmetric SQUID; for a full treatment including finite junction asymmetry see Refs.~\cite{kounalakis2022analog,kounalakis2020flux}.
Replacing $E_J$ with $E_J^\prime$ in Eq.~(\ref{eq:CircuitHam}) and expressing the transmon operators in terms of annihilation (creation) operators $\hat{c}^{(\dagger)}$, i.e., ${\hat{N}=i\left[E_{J}/(32E_{C})\right]^{1/4}(\hat{c}^{\dagger}-\hat{c})}$, ${\hat{\delta}=\left[2E_{C}/E_{J}\right]^{1/4}(\hat{c}+\hat{c}^{\dagger})}$~\cite{vool2017introduction}, yields the total system Hamiltonian
\begin{equation}
\hat{H} =\hat{H}_{q} +  \hbar\sum_j\left[\omega_{j}\hat{a}_j^{\dagger}\hat{a}_j -  g_j \hat{c}^{\dagger}\hat{c}(\hat{a}_j+\hat{a}_j^{\dagger})\right],
\label{eq:HamTotal}
\end{equation}
where $\hat{H}_{q}=\hbar\omega_{q}\hat{c}^{\dagger}\hat{c}-\frac{E_{C}}{2 }\hat{c}^{\dagger}\hat{c}^{\dagger}\hat{c}\hat{c}$, is the bare transmon Hamiltonian (valid for ${E_J\gg E_{C}}$), with qubit frequency $\omega_{q}=(\sqrt{8E_JE_{C}}-E_{C})/\hbar$~\cite{koch2007charge}.

The last term in Eq.~(\ref{eq:HamTotal}) describes the radiation-pressure interaction between the qubit and each bosonic mode, with coupling strength
\begin{equation}
g_j= \frac{\partial\omega_q}{\partial\phi_j}\phi_j^\mathrm{zpf},
\label{eq:couplings}
\end{equation}
where $\phi_j^\mathrm{zpf}$ is the magnitude of the flux fluctuations induced by either the beam or the magnet, given by $\phi_x^\mathrm{zpf}=\pi\beta_0B_zlx_\mathrm{zpf}/\Phi_0$ and $\phi_m^\mathrm{zpf}=\mu_0\mu_\mathrm{zpf}/(4\sqrt{2}d\Phi_0)$, respectively.
In the case of a symmetric SQUID, the transmon frequency sensitivity to flux changes is 
\begin{equation}
\frac{\partial\omega_q}{\partial\phi_j}=\frac{\omega_p}{2}\frac{\sin{\phi_b}}{\sqrt{\cos{\phi_b}}},  
\end{equation}
where $\omega_{p}=\sqrt{8E_{J}^{\mathrm{max}}E_{C}}/\hbar$ is the Josephson plasma frequency at $\phi_b=2\pi k~(k\in \mathbb{Z})$.
The behavior of the coupling strength as a function of the SQUID asymmetry and $\phi_b$ is studied in detail in Ref.~\cite{kounalakis2022analog}.

\section{Entangled coherent state generation}

The system Hamiltonian in Eq.~(\ref{eq:HamTotal}) describes a qubit interacting with a set of bosonic modes via bipartite radiation-pressure interactions.
However, in the absence of additional driving, these radiation-pressure couplings lead to interesting dynamics only in the \emph{ultrastrong} coupling regime, ${g_j\gtrsim\omega_j}$~\cite{nunnenkamp2011single,nation2016ultrastrong,kounalakis2020flux}.
Typically, mechanical beam resonators have frequencies of a few MHz~\cite{rodrigues2019coupling,schmidt2020sideband} and operating magnon frequencies lie above 100~MHz~\cite{tabuchi2016quantum,rameshti2021cavity}, whereas $g_j\lesssim10~\mathrm{MHz}$~\cite{kounalakis2020flux,kounalakis2022analog}.
Therefore, while the ultrastrong coupling condition seems promising for optomechanical setups~\cite{kounalakis2020flux}, it is far from realistic for magnonic devices.
On the other hand, when external driving is introduced to the system, the radiation-pressure interaction can be ``activated'' even for ${g_j<\omega_j}$, e.g., by a stroboscopic application of short $\pi$ qubit pulses~\cite{tian2005entanglement} or by modulating the coupling~\cite{kielpinski2012quantum,kounalakis2022analog}.

Here, without loss of generality, we consider the case ${g_j\ll\omega_j}$ and assume that the radiation-pressure interaction is activated by applying a weak flux modulation through the SQUID loop as in Ref.~\cite{kounalakis2022analog}.
In this scheme the qubit operates around the transmon ``sweetspot'', i.e., $\phi_b\simeq0$, and an applied ac flux with amplitude $\phi_\mathrm{ac}$ at frequency $\omega_\mathrm{ac}$, modulates the flux, $\phi_b=\phi_\mathrm{ac}\cos{(\omega_\mathrm{ac}t-\theta)}\ll1$, resulting in a modulated coupling strength $g_j(t)=\frac{\omega_p}{2}\phi_j^\mathrm{zpf}\cos{(\omega_\mathrm{ac}t-\theta)}$, where $\theta$ is a constant phase.
In the frame rotating at $\omega_\mathrm{ac}$ the transformed Hamiltonian reads,
\begin{align}
\hat{\widetilde{H}} = \hat{H}_{q} + \hbar\sum_j\left[\Delta_{j}\hat{a}_j^{\dagger}\hat{a}_j -  \widetilde{g}_j \hat{c}^{\dagger}\hat{c}(\hat{a}_je^{i\theta}+\hat{a}_j^{\dagger}e^{-i\theta})\right],
\label{eq:HamTotaltilde}
\end{align}
where $\widetilde{g}_j=\frac{\omega_p}{4}\phi_j^\mathrm{zpf}$, $\Delta_{j}=\omega_{j}-\omega_\mathrm{ac}$ and we have omitted fast-rotating terms $\hat{c}^{\dagger}\hat{c}\hat{a}_j^{(\dagger)}e^{\pm i(\omega_j+\omega_\mathrm{ac}) t}$ which do not contribute to the dynamics since $\widetilde{g}_i\ll(\omega_j+\omega_\mathrm{ac})$.

We now describe a simple protocol for generating ECS that are maximally entangled using the Hamiltonian in Eq.~(\ref{eq:HamTotaltilde}).
Let us assume there are $N$ bosonic modes, interacting with the qubit via bipartite radiation-pressure couplings.
First, a microwave pulse, prepares the qubit in a superposition state ${\ket{\chi}_q\dot{=}(\ket{0_{q}}+e^{i\chi}\ket{1_{q}})/\sqrt{2}}$.
The next step is to activate the bipartite interaction of the qubit with each mode.
In the simple case where all the modes we want to entangle have the same frequency, $\omega_j$, then by turning on the flux modulation, i.e., setting $\omega_\mathrm{ac}=\omega_j$, for a variable duration, $\tau_j$, the system evolves into a hybrid generalized Greenberger--Horne--Zeilinger state
\begin{equation}
\ket{\psi}_\mathrm{GHZ}=\frac{1}{\mathcal{N}}\left(\ket{0_{q} 0_{1} \cdots 0_{N}} + e^{i\chi} \ket{1_{q}\alpha_{1} \cdots \alpha_{N}}\right),
\label{GHZhybrid}
\end{equation}
where $\ket{\alpha_j}$ denotes a coherent state with complex phase space amplitude $\alpha_j=-i\widetilde{g}_j\tau_j$.
For $\abs{\alpha_j}\gtrsim4$ the normalization factor is $\mathcal{N}\simeq\sqrt{2}$~\cite{mirrahimi2014dynamically}.
Note that if there are $M$ modes with different frequencies, then the flux modulation should be activated $M$ times in order to prepare the state in Eq.~(\ref{GHZhybrid}).
% Experimentally, this operation can be performed via the same flux bias channel by appropriately adjusting the modulation times and frequencies.

Applying a qubit pulse $R_{\hat{y},\frac{\pi}{2}}$ followed by a strong projective measurement collapses the qubit in its ground or excited state and projects the bosonic system into $\frac{1}{\mathcal{N}_\pm}\left(\ket{0_10_2\cdots0_N} \pm e^{i\chi}\ket{\alpha_{1}\alpha_{2} \cdots \alpha_{N}}\right)$, where the ``$+$'' or ``$-$'' state results from measuring the qubit in $\ket{0_q}$ or $\ket{1_q}$, respectively.
For the case of two bosonic modes with $\widetilde{g}_{1,2}, \tau_{1,2}$ chosen such that $\alpha_1=\alpha_2=\alpha$ and $\chi=0$ the prepared state corresponds to the maximally-entangled \emph{Bell} state, 
\begin{equation}
\ket{\pm\Psi_\mathrm{Bell}}=\frac{1}{\mathcal{N}_\pm}\left(\ket{0 0} \pm \ket{\alpha\alpha}\right),
\label{Psi_Bell}
\end{equation}
where $\mathcal{N}_\pm=\sqrt{2(1\pm e^{-|\alpha|^2})}\simeq\sqrt{2}$ for $\abs{\alpha}\gtrsim4$~\cite{ralph2003quantum}.
Alternatively, in the case of different frequency modes, $\omega_1\neq\omega_2$, a maximally-entangled \emph{NOON} state of the form 
\begin{equation}
\ket{\pm\Phi_\mathrm{NOON}}=\frac{1}{\mathcal{N}_\pm}\left(\ket{0 \alpha} \pm \ket{\alpha 0}\right),
\end{equation}
can be obtained by performing a $\pi$ pulse to flip the qubit state right after turning on the first interaction and before the second one.
The protocol would then require the following steps: (a) start modulating at $\omega_\mathrm{ac}=\omega_1$, (b) turn off the interaction after time $\tau_1$, (c) apply $\pi$ qubit pulse, and (d) switch on the second flux modulation with $\omega_\mathrm{ac}=\omega_2$ for time $\tau_2=\tau_1\widetilde{g}_1/\widetilde{g}_2$.

Additionally, more general ECS of the form,
\begin{equation}
\ket{\Psi}_\text{ij} = c_{00} \ket{0_{i} 0_{j} } + c_{1\alpha} \ket{0_{i} \alpha_{j} } + c_{\alpha0} \ket{\alpha_{i} 0_{j}} + c_{\alpha\alpha} \ket{\alpha_{i}\alpha_{j}},
\label{Psi_general}
\end{equation}
with $c_{\alpha0},c_{0\alpha}\neq0$, may also be generated using appropriately adjusted protocols.
For example, starting from $\ket{\psi}_\mathrm{qij}=\ket{(0_q+1_q)0_i0_j}$, then turning on the interaction with mode $i$ for time $\tau_i$ such that $\abs{\alpha}\equiv\abs{\widetilde{g}_i\tau_i}\gtrsim4$, and applying a $R_{\hat{y}\frac{\pi}{2}}$ qubit pulse, results in the state $\ket{\psi}_\mathrm{qij}=\frac{1}{2}\left[\ket{0_q0_i0_j}+\ket{0_q\alpha_i0_j}+\ket{1_q0_i0_j}-\ket{1_q\alpha_i0_j}\right]$.
If we subsequently turn on the interaction with mode $j$ (for time $\tau_j=\alpha/\widetilde{g}_j$) and apply another $R_{\hat{y}\frac{\pi}{2}}$ qubit pulse, the resulting state is, $\ket{\psi}_\mathrm{qij}=\frac{1}{\sqrt{2}}\left(\ket{0}_q\ket{\Psi}_\text{ij}^+ +\ket{1}_q\ket{\Psi}_\text{ij}^-\right)$, where
\begin{equation}
\ket{\Psi}_\text{ij}^\pm=\frac{1}{2}\left(\ket{0_{i} 0_{j} } +  \ket{0_{i} \alpha_{j} } \pm \ket{\alpha_{i} 0_{j}} \mp \ket{\alpha_{i}\alpha_{j}}\right).
\end{equation}
Finally, a strong measurement collapses the qubit in $\ket{0}_q$ or $\ket{1}_q$, projecting the system in the maximally-entangled two-mode state $\ket{\Psi}_\text{ij}^+$ or $\ket{\Psi}_\text{ij}^-$, respectively.

\section{Numerical modeling \& benchmarking}

We benchmark the protocol described above for generating the Bell state $\ket{+\Psi_\mathrm{Bell}}$ against realistic experimental conditions including dissipation using the quantum statistical Lindblad master equation~\cite{johansson2012qutip}
\begin{align}
\dot{\rho}=&\frac{i}{\hbar}[\rho,\hat{\widetilde{H}}]+\sum_j\frac{\omega_{j}}{Q_j}\left(n_j^\mathrm{th}\mathcal{L}[\hat{a}_j^\dagger]\rho+(n_\mathrm{th}+1)\mathcal{L}[\hat{a}_j]\rho\right)\nonumber\\
&+\frac{1}{T_1}\mathcal{L}[\hat{c}]\rho+\frac{1}{T_2}\mathcal{L}[\hat{c}^\dagger\hat{c}]\rho,
\label{eq:Lindblad}
\end{align}
where $Q_j$ is the quality factor of each resonator, $\mathcal{L}[\hat{o}]\rho=(2\hat{o}\rho\hat{o}^\dagger-\hat{o}^\dagger\hat{o}\rho-\rho\hat{o}^\dagger\hat{o})/2$ are superoperators describing each bare dissipation channel and ${n_j^\mathrm{th}=1/[\exp(\hbar\omega_j/(k_BT))-1]}$ is the number of thermally excited magnons/phonons at temperature $T$.
$T_1$ and $T_2$ are the qubit relaxation and dephasing times, respectively, for which we pick a realistic value of $50~\mathrm{\mu s}$ throughout our simulations~\cite{kjaergaard2020superconducting}.
Of note, the in-plane magnetic field that is required to enable the qubit coupling to the magnonic or the mechanical resonator, $B_z\sim10-50~\mathrm{mT}$~\cite{kounalakis2020flux,kounalakis2022analog}, is not expected to limit the qubit performance~\cite{krause2021magnetic}.
In addition, while the transmon is effectively a qubit, it is more accurately described as a three-level system with negative anharmonicity given by $\sim-E_C$.
We therefore model it as such choosing a typical value of $E_C/h=300$~MHz~\cite{koch2007charge,kjaergaard2020superconducting}.

\begin{figure}[t]
  \begin{center}
  \includegraphics[width=1.0\linewidth]{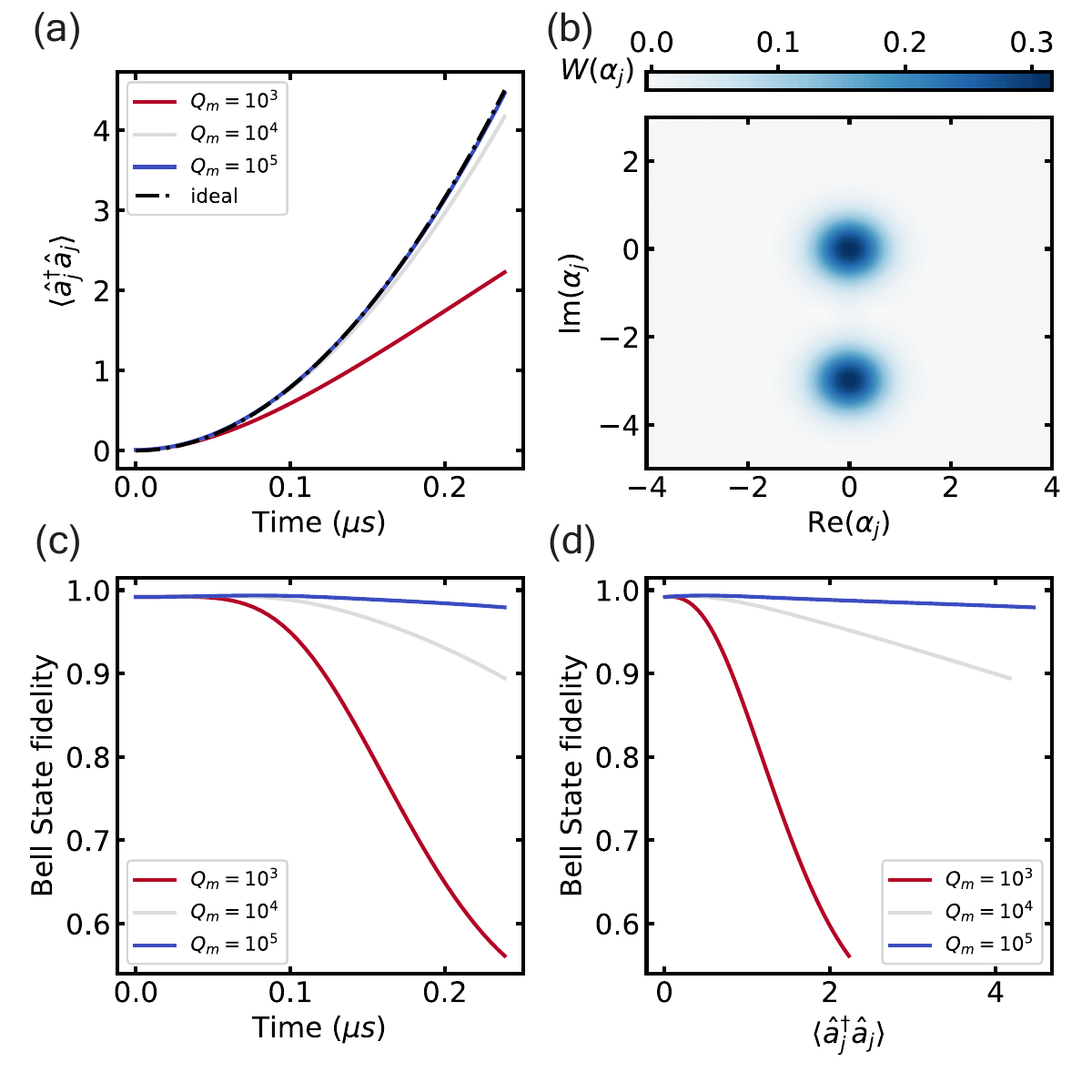
  }
  \end{center}
  \caption{
  {Bell state benchmarking for the case of two Kittel modes in two identical YIG spheres, as schematically shown in Fig.~\ref{fig:scheme}(a).
  (a)~Magnon number in each magnonic mode, as a function of time during the protocol, shown for different resonator quality factors.
  (b)~Wigner function of the individual magnonic state in one mode, after tracing out the other mode, at the end of the protocol for $Q_m=10^5$.
  The fidelity of the prepared state to the ideal Bell state $\ket{+\Psi_\mathrm{Bell}}$ is shown as a function of time in (c) and as a function of the magnon number in (d).
  System parameters: $\omega_{1,2}/(2\pi)=1~\mathrm{GHz}$,  $\widetilde{g}_{1,2}/(2\pi)=2~\mathrm{MHz}$, $T_1=T_2=50~\mu\mathrm{s}$, $T=10~\mathrm{mK}$.}
  }
  \label{fig:BellState}
\end{figure}

We first study the case, schematically depicted in Fig.~\ref{fig:scheme}(a), of two YIG spheres placed diametrically opposite with respect to the center of the SQUID.
For simplicity, we assume two identical spheres and Kittel modes with the same frequency, $\omega_{1,2}/(2\pi)=1~\mathrm{GHz}$, as well as coupling to the qubit, $\widetilde{g}_{1,2}/(2\pi)=2~\mathrm{MHz}$, and study the performance of the protocol proposed above as a function of the resonator quality factor, $Q_m$, at $T=10~\mathrm{mK}$ ($n^\mathrm{th}_{1,2}\simeq0.01$).
For typical values of the Gilbert damping constant $\alpha_{G}$ we expect $Q_m=1/\alpha_{G}\sim10^{3}-10^{5}$~\cite{tabuchi2015coherent,klingler2017gilbert,rameshti2021cavity}.

In Fig.~\ref{fig:BellState}(a) we plot the evolution of the magnon number in either mode $j$ and compare it to the ideal case, i.e., without dissipation, where $\langle \hat{a}_j^\dagger\hat{a}_j\rangle(t)=\abs{\widetilde{g}_mt}^2/2$.
In addition, in Fig.~\ref{fig:BellState}(b) we plot the  Wigner quasi-probability distribution at $t=0.24~\mu\mathrm{s}$ for $Q_m=10^5$, which is defined as ${W(\alpha_j)=2/\pi\Tr{D^\dagger(\alpha_j)\rho_j D(\alpha_j)e^{i\pi \hat{a}_j^\dagger\hat{a}_j}}}$, where $\rho_j\equiv\Tr_i{\left[\rho_{ij}\right]}$ is the reduced density matrix of mode $j$ and ${D(\alpha_j)=e^{\alpha\hat{a}_j^\dagger-\alpha^{*}\hat{a}_j}}$ is the displacement operator acting on this mode.
The two-mode density matrix, $\rho_{ij}$, is obtained after projecting on $\ket{+_q}$, and tracing out the qubit, i.e., $\rho_{ij}\equiv\Tr_q{\left[\rho\ket{+_q} \bra{+_q}\right]}$.
We note that since we have two identical modes, the magnon number evolution as well as the reduced-state Wigner functions are exactly the same for both.
Furthermore, Figs.~\ref{fig:BellState}(c) and \ref{fig:BellState}(d) show the fidelity $\mathcal{F}=\sqrt{\langle+\Psi_\mathrm{Bell}\vert\rho_{12}\ket{+\Psi_\mathrm{Bell}}}$~\cite{nielsen2010quantum,johansson2012qutip} of the prepared two-mode state to the ideal Bell state, as a function of time and magnon number, respectively.
Evidently, for realistic values of the magnonic quality factors $Q_m\gtrsim10^{4}$~\cite{tabuchi2015coherent,klingler2017gilbert,khosla2018displacemon}, the desired Bell state can be prepared with high fidelity $\mathcal{F}\lesssim90\%$.

\begin{figure}[t]
  \begin{center}
  \includegraphics[width=1.0\linewidth]{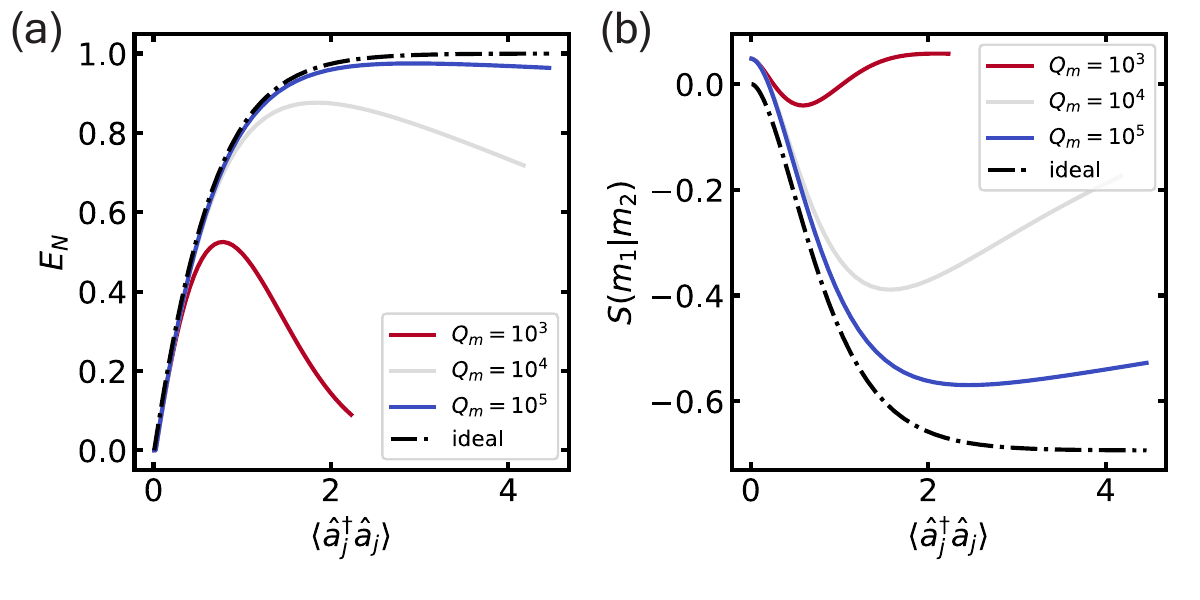}
  \end{center}
  \caption{
  (a)~Logarithmic negativity and (b)~conditional quantum entropy as a function of the magnon number for the magnon-magnon system described in Fig.~\ref{fig:BellState}.
  In the absence of dissipation (dashed-dotted curves) an ideal Bell state is created for magnon numbers $\langle \hat{a}_j^\dagger\hat{a}_j\rangle>2$ with $E_N\rightarrow1$ and $S(m_1|m_2)\rightarrow-\log2$.
  }
  \label{fig:BellStateEntanglement}
\end{figure}

To showcase the evolution of the bipartite entanglement during the protocol, in Fig.~\ref{fig:BellStateEntanglement}(a) we plot the logarithmic negativity $E_N=\log_2(2N(\rho_{12})+1)$, where $N(\rho_{12})$ is the sum of negative eigenvalues of the partial transpose of the two-mode density matrix $\rho_{12}$~\cite{vidal2002computable}.
The dashed-dotted curve shows the logarithmic negativity evolution in the ideal case, $E_N(t)=\log_2\left[2/(e^{-\abs{\widetilde{g}_{j}t}^2}+1)\right]$~\cite{liu2014quantum}.
For $\abs{\alpha}\equiv\abs{\widetilde{g}_{j}t}\gtrsim2$ it approaches the ideal value of $E_N^\mathrm{max}=1$, where the two modes are maximally entangled, before magnon dissipation eventually takes over and the entanglement gets lost.

Furthermore, in Fig.~\ref{fig:BellStateEntanglement}(b) we plot the conditional quantum entropy $S(m_1|m_2)=S(\rho_{12})-S(\rho_2)$~\cite{cerf1997negative,horodecki2005partial}, where $S(\rho_{ij})$ and $S(\rho_j)$ are the Von Neumann entropies of the joint and reduced state, respectively, with $S(\rho)=-\Tr[\rho\ln{\rho}]$.
Negative conditional quantum entropy serves as a sufficient criterion for the quantum state to be entangled and provides a measure of the degree of coherent quantum communication between the two entangled modes~\cite{cerf1997negative,horodecki2005partial}.
For maximally-entangled Bell states we have $S(\rho_{ij})=0$ and $S(\rho_{j})=\ln{2}$.
Therefore, in the limit of large magnon numbers, we expect $S(m_1|m_2)\rightarrow-\ln{2}$, as illustrated by the dashed-dotted curve plotting the ideal (dissipationless) case.
However, as the entanglement starts decreasing due to magnon dissipation, the joint entropy of the system becomes positive and both $S(\rho_{ij})$ and $S(\rho_{i})$ start increasing.
Therefore, as expected, the positive value threshold for $S(m_1|m_2)$ is surpassed faster and at lower magnon numbers as the quality factors get smaller.
Note that initially $S(m_1|m_2)>0$ due to the fact that the modes start in a thermal state with $n_\mathrm{th}\simeq0.01$.

\begin{figure}[t]
  \begin{center}
  \includegraphics[width=1.0\linewidth]{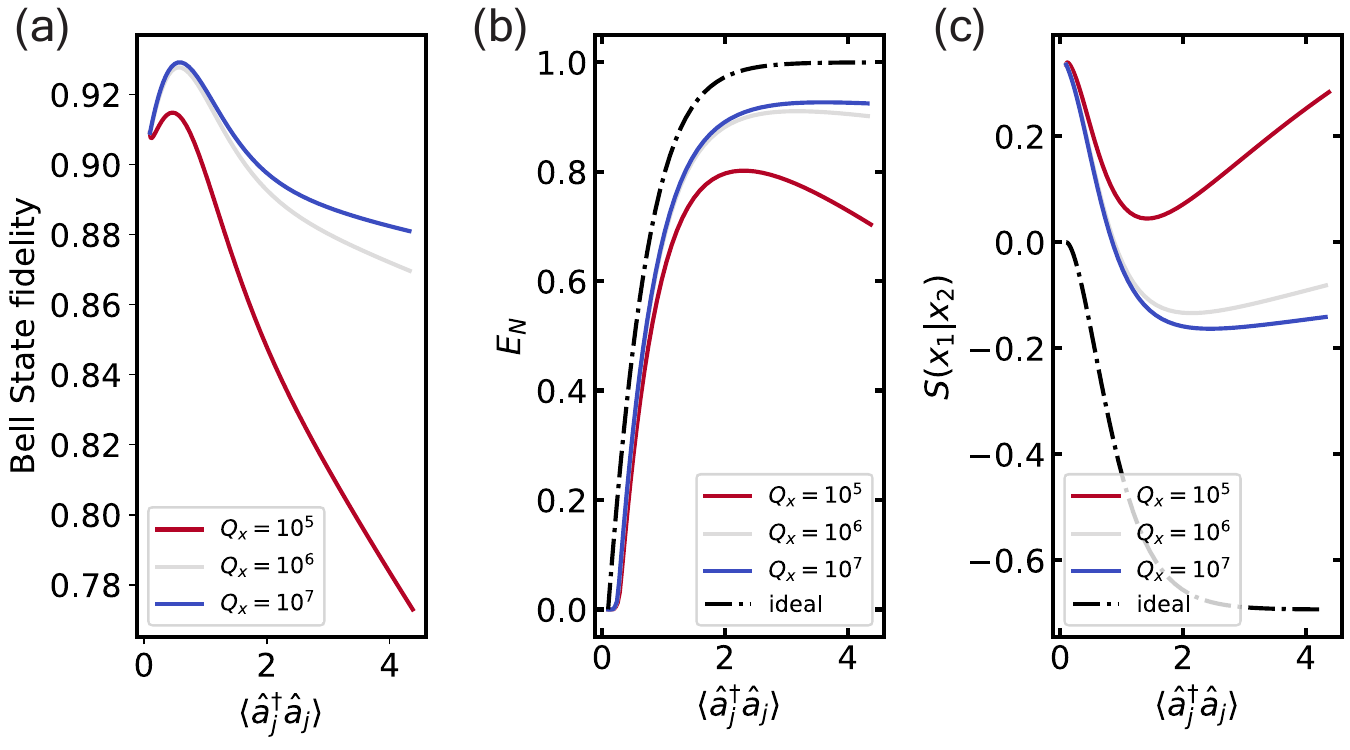}
  \end{center}
  \caption{
  {(a)~Bell state fidelity, (b)~logarithmic negativity and (c)~conditional quantum entropy as a function of the phonon number for the case of two SQUID-embedded mechanical nanobeams interacting via the transmon.
  System parameters: $\omega_{1,2}/(2\pi)=10~\mathrm{MHz}$,  $\widetilde{g}_{1,2}/(2\pi)=100~\mathrm{kHz}$, $T_1=T_2=50~\mu\mathrm{s}$, $T=10~\mathrm{mK}$, initial $n_{1,2}^\mathrm{th}=0.1$.}
  }
  \label{fig:PhononBellStateEntanglement}
\end{figure}

The protocol described above can also be applied to entangle mechanical beam resonators embedded in the SQUID loop, as depicted in Fig.~\ref{fig:scheme}(b).
These can be realized using carbon nanotubes~\cite{khosla2018displacemon} or aluminum-based mechanical beams~\cite{rodrigues2019coupling,schmidt2020sideband,bera2021large,kounalakis2020flux} interacting via radiation-pressure couplings with the transmon.
The former have operating frequencies and quality factors similar to the magnonic case studied above, therefore, the results in Figs.~\ref{fig:BellState} and \ref{fig:BellStateEntanglement} are applicable as well.
On the other hand, mechanical beam resonators made of aluminum typically operate in the range $1-10$~MHz, with quality factors $Q_x\gtrsim10^5$~\cite{rodrigues2019coupling,schmidt2020sideband,bera2021large}.

Therefore, in conjunction with the magnonic case, we numerically test the same protocol for creating mechanical Bell states between two SQUID-embedded aluminum beam resonator modes~\cite{schmidt2020sideband}, with the same frequency $\omega_{1,2}/(2\pi)=10$~MHz and coupling to the qubit $\widetilde{g}_{1,2}/(2\pi)=100$~kHz.
Typical temperatures of $T\sim10~\mathrm{mK}$, correspond to high thermal population at these frequencies, however, cooling schemes can reduce the number of thermal phonons to $\lesssim0.1$~\cite{kounalakis2020flux,khosla2018displacemon}.
We therefore assume an attainable initial thermal population $n_{1,2}^\mathrm{th}=0.1$ and an operating temperature of $T=10~\mathrm{mK}$.

In Figs.~\ref{fig:PhononBellStateEntanglement}(a) and \ref{fig:PhononBellStateEntanglement}(b) we plot the Bell-state fidelity and the logarithmic negativity, respectively, as a function of the phonon number during the protocol for quality factors in the range $Q_x=10^5-10^7$.
Note that initially the fidelity is less than 1, due to the finite thermal population in both resonators, however, as the protocol evolves it starts increasing before phonon dissipation takes over.
We find that, for realistic quality factors $Q_x\gtrsim10^6$, high phonon number Bell-states can be prepared with high fidelity and sufficiently high entanglement as quantified by $E_N$.
However, as shown in Fig.~\ref{fig:PhononBellStateEntanglement}(c), the effects of the initial thermal population seem to be detrimental to the conditional quantum entropy $S(x_1|x_2)$ which remains far from the ideal limit during the whole protocol and only reaches negative values for $Q_j\sim10^6$.

Experimental verification of the prepared states can be obtained by performing state tomography.
For example, in the case of mechanical resonators, by sideband driving on the qubit one may engineer beam-splitter and two-mode squeezing interactions that can be used to detect correlations of the entangled state similar to Ref.~\cite{kotler2021direct}.
This method may also be applied to the magnonic resonators, for which independent state tomography techniques exist as well~\cite{hioki2021state}.
However, strong driving may severely impact the qubit state~\cite{lescanne2019observing} limiting the success of such protocols.
For this reason we have also analyzed an alternative scheme for reading out the entangled states, presented in the Appendix, which relies solely on switching on/off the interaction and performing magnon/phonon displacements and qubit measurements.

\section{Conclusion}
In summary, we have proposed a scheme for generating ECS of magnons/phonons in a hybrid circuit QED architecture comprising a superconducting transmon qubit coupled to different magnonic/mechanical modes via bipartite flux-mediated interactions.
In particular, we have highlighted several schemes for creating maximally-entangled states and, as a proof-of-principle demonstration, we have numerically tested a simple protocol for generating magnonic and mechanical Bell states under realistic experimental conditions.
We show that high-fidelity Bell states can be prepared in the presence of typical dissipation mechanisms in the system.
Furthermore, in the Appendix we have analyzed a readout scheme, using standard circuit operations, that can be used as an alternative to existing tomography methods for verifying the prepared states.
Our results pave the way towards creating controllable quantum networks of entangled magnons in a flexible and scalable platform without relying on microwave 3D cavities or strong driving.
Although for simplicity we have considered identical YIG spheres, our results are also applicable to nonidentical modes and other geometries such as micro-disk resonators~\cite{srivastava2023identification}.
Finally, as we demonstrate numerically, the proposed scheme for creating and controlling ECS is also applicable to SQUID-embedded mechanical beam resonators, opening up new opportunities for quantum information tasks in this platform and potentially giving rise to novel magnonic-mechanical hybrid devices.\\

% The magnitude of both couplings can be tuned via a DC flux bias $\phi_b$, however, in the case of identical spheres and equal distance from the SQUID center the two coupling strengths are always the same.
% While Eq.~(\ref{eq:couplings}) suggests that $g_i$ is infinite at $\phi_b=(k+\frac{1}{2})\pi$, we note that a finite junction asymmetry shifts the optimum bias point and results in a finite maximum coupling~\cite{shevchuk2017strong,kounalakis2019synthesizing,kounalakis2022analog}.

% In addition, we do not externally drive the transmon beyond the single-excitation manifold and the available interactions do not cause level transitions, therefore we can treat it as a qubit and express the total system Hamiltonian as
% \section{Phonon-phonon \& magnon-phonon entanglement}

% We note that our methods also apply to the case of coupling two bosonic modes in SQUID-embedded mechanical resonators~\cite{schmidt2020sideband,kounalakis2019synthesizing,kounalakis2020flux,bera2021large}, or between such a mechanical resonator and YIG sphere.

\section*{Acknowledgments}
We thank Sanchar Sharma and Victor Bittencourt for helpful discussions.
This research was supported by the Dutch Foundation for Scientific Research (NWO).
M.K. and S.V.K. would like to acknowledge financial support by the German Federal Ministry of Education and Research (BMBF) project QECHQS (Grant No. 16KIS1590K).\\

\section*{Appendix: Readout scheme}

\renewcommand{\thesection}{A\arabic{section}}
\renewcommand{\theequation}{A\arabic{equation}}
\renewcommand{\thefigure}{A\arabic{figure}}
\renewcommand{\thetable}{A\arabic{table}}

We now describe a method for reading out the two-mode ECS discussed in the main text, using only qubit measurements and displacement operations on the bosonic modes.
We start with the assumption that the most general state one can prepare with the system Hamiltonian in Eq.~(\ref{eq:HamTotaltilde}) is of the following form,
\begin{align}
\ket{\Psi}_\text{ij}=&c_0 e^{i\theta_0} \ket{0_{i} 0_{j} } + c_1 e^{i\theta_1} \ket{0_{i} \alpha_{j} } \nonumber\\
&+ c_2 e^{i\theta_2} \ket{\alpha_{i} 0_{j}} + c_3 e^{i\theta_3} \ket{\alpha_{i}\alpha_{j}},
\label{eq:Psi_general_SI}
\end{align}
where $c_j$ are real positive numbers and $\sum_{j=0}^{3}c_j^2=1$.
Our assumption is based on the fact that the engineered radiation-pressure interaction in Eq.~(\ref{eq:HamTotaltilde}) can only lead to magnon/phonon displacements when the qubit is in the excited state, therefore, for the protocols described in the main text, where the interaction is activated at least once for each bosonic mode, Eq.~(\ref{eq:Psi_general_SI}) describes the mode general state one can prepare.
In addition, single-photon losses acting on coherent states result in a coherent state of smaller amplitude, therefore this decay channel does not alter the form of the state described in Eq.~(\ref{eq:Psi_general_SI}).

Assuming the state in Eq.~(\ref{eq:Psi_general_SI}) has been prepared, we start the readout protocol by preparing the qubit in a general superposition state $\ket{\phi}_\text{q}=(\ket{0}_\text{q}+e^{i\phi}\ket{1}_\text{q})/\sqrt{2}$.
After switching on both interactions, the system wavefunction evolves as
\begin{widetext}
\begin{align}
U_\mathrm{int}^{(i)}&U_\mathrm{int}^{(j)}\ket{\phi}_\text{q}\ket{\Psi}_\text{ij} = \frac{1}{\sqrt{2}}\Bigl[\ket{0}_\text{q}\left(c_0 e^{i\theta_0} \ket{0_{i}0_{j}} + c_1 e^{i\theta_1} \ket{0_{i}\alpha_{j}} + c_2 e^{i\theta_2} \ket{\alpha_{i} 0_{j}} + c_3 e^{i\theta_3} \ket{\alpha_{i}\alpha_{j}}\right)+ \nonumber\\
&\ket{1}_\text{q}e^{i(\phi+\bar{\phi})}\left(c_0 e^{i\theta_0} \ket{\beta_{i}\beta_{j}} + c_1 e^{i\theta_1+\gamma_{j}} \ket{\beta_{i}(\alpha+\beta)_{j}} + c_2 e^{i\theta_2+\gamma_{i}} \ket{(\alpha+\beta)_{i} \beta_{j}} + c_3 e^{i\theta_3+\gamma_{i}+\gamma_{j}} \ket{(\alpha+\beta)_{i}(\alpha+\beta)_{j}}\right)\Bigr],
\end{align}
\end{widetext}
where $U_\mathrm{int}^{(j)}=\exp{i  \widetilde{g}_j \hat{c}^{\dagger}\hat{c}(\hat{a}_je^{i\theta}+\hat{a}_j^{\dagger}e^{-i\theta})t}$.

The displacement amplitudes and corresponding geometric phases, which arise from the radiation pressure interactions, are given by $\beta_\text{i,j}~\dot{=}~\beta(t_\text{i,j})=(g_\text{i,j}/\omega_\text{i,j})~(e^{i\omega_\text{i,j}t_\text{i,j}}~-~1)$ and $\bar{\phi}~\dot{=}~\bar{\phi}(t_\text{i,j})=(g_\text{i,j}/\omega_\text{i,j})^2~(\omega_\text{i,j}t_\text{i,j}~-~\sin{(\omega_\text{i,j}t_\text{i,j})})$~\cite{asadian2014probing, kounalakis2022analog}.
For simplification purposes we have assumed that the latter are equal and, since $\phi$ is arbitrarily determined at the qubit preparation stage, they can be absorbed into a redefinition of $\phi\rightarrow \bar{\phi}+\phi$.
The phases $\gamma_\text{i,j}=\text{Im}(\alpha^*\beta_\text{i,j})$ arise from the fact that in general two consecutive displacements do not commute.

The above state can also be written as
\begin{equation}
\ket{\psi}_\text{qij} = \frac{1}{2}\left(\ket{+}_\text{q} \ket{\Psi^+}_\text{ij}+\ket{-}_\text{q} \ket{\Psi^-}_\text{ij}\right),
\end{equation}
where $\ket{\pm}=(\ket{0}\pm\ket{1})/\sqrt{2}$ are the eigenstates of the Pauli $\hat{\sigma}_x$ operator and
\begin{widetext}
\begin{align}
\ket{\Psi^\pm}_\text{ij}&=c_0 e^{i\theta_0} \ket{0_{i}0_{j}} + c_1 e^{i\theta_1} \ket{0_{i}\alpha_{j}} + c_2 e^{i\theta_2} \ket{\alpha_{i} 0_{j}} + c_3 e^{i\theta_3} \ket{\alpha_{i}\alpha_{j}}\nonumber \\ 
&\pm\Bigl(c_0 e^{i\theta_0+\phi} \ket{\beta_{i}\beta_{j}} + c_1 e^{i\theta_1+\phi+\gamma} \ket{\beta_{i}(\alpha+\beta)_{j}} +c_2 e^{i\theta_2+\phi+\gamma} \ket{(\alpha+\beta)_{i} \beta_{j}} + c_3 e^{i\theta_3+\phi+2\gamma} \ket{(\alpha+\beta)_{i}(\alpha+\beta)_{j}}\Bigr).
\label{eq:Psi_displaced}
\end{align}
\end{widetext}
The expectation value of the qubit in the $\ket{\pm}$ basis is then given by
\begin{equation}
\langle \hat{\sigma}_x \rangle_{\beta,\beta}=\frac{1}{4}\left(|\langle {\Psi^+}_\text{ij}|{\Psi^+}_\text{ij} \rangle|^2-|\langle {\Psi^-}_\text{ij}|{\Psi^-}_\text{ij} \rangle|^2\right).
\label{eq:X_mean_general}
\end{equation}
We now consider several cases for each displacement:

(I) First, assuming the coupling strength and interaction times for both resonators are chosen such that $\beta_\text{i,j}=\alpha_\text{i,j}$, we have ($\gamma_\text{i,j}=0$):
\begin{widetext}
\begin{align}
\ket{\Psi^\pm}_\text{ij}=c_0 e^{i\theta_0} \ket{0_{i}0_{j}} + c_1 e^{i\theta_1} \ket{0_{i}\alpha_{j}} + c_2 e^{i\theta_2} \ket{\alpha_{i} 0_{j}} + &(c_3 e^{i\theta_3} \pm c_0 e^{i\theta_0+\phi}) \ket{\alpha_{i}\alpha_{j}} \pm  c_1 e^{i\theta_1+\phi} \ket{\alpha_{i} (2\alpha)_{j}}\nonumber \\
& \pm   c_2 e^{i\theta_2+\phi} \ket{(2\alpha)_{i} \alpha_{j}}\pm c_3 e^{i\theta_3+\phi} \ket{(2\alpha)_{i} (2\alpha)_{j}}
\label{eq:Psi_displaced_aa}
\end{align}
\end{widetext}
From Eq.~(\ref{eq:X_mean_general}) we obtain
\begin{align}
\langle \hat{\sigma}_x \rangle_{\alpha,\alpha}~&=~|c_3 e^{i\theta_3} + c_0 e^{i\theta_0+\phi}|^2-|c_3 e^{i\theta_3} - c_0 e^{i\theta_0+\phi}|^2\nonumber \\ 
&=c_0c_3~\cos{(\phi+\theta_0-\theta_3)}.
\label{eq:X_mean_++}
\end{align}
Additionally, for $\beta_\text{i,j}=-\alpha_\text{i,j}$ it can be shown that 
\begin{equation}
\langle \hat{\sigma}_x \rangle_{-\alpha,-\alpha}=c_0c_3~\cos{(\phi+\theta_3-\theta_0)}.
\label{eq:X_mean_--}
\end{equation}

(II) For the case $\beta_{i}=\alpha_{i}$, $\beta_{j}=-\alpha_{j}$, using Eq.~(\ref{eq:Psi_displaced}) and Eq.~(\ref{eq:X_mean_general}), it follows that
\begin{equation}
\langle \hat{\sigma}_x \rangle_{\alpha,-\alpha}=c_1c_2~\cos{(\phi+\theta_1-\theta_2)}.
\label{eq:X_mean_+-}
\end{equation}
Similarly for $\beta_{i}=-\alpha_{i}$, $\beta_{j}=\alpha_{j}$ we obtain
\begin{equation}
\langle \hat{\sigma}_x \rangle_{-\alpha,\alpha}=c_1c_2~\cos{(\phi+\theta_2-\theta_1)}
\label{eq:X_mean_-+}
\end{equation}

(III) For the cases $\beta_{i}=\alpha_{i}$, $\beta_{j}=0$ and $\beta_{i}=-\alpha_{i}$, $\beta_{j}=0$ we have
\begin{equation}
\langle \hat{\sigma}_x \rangle_{\alpha,0}=c_0c_2~\cos{(\phi+\theta_0-\theta_2)}+c_1c_3~\cos{(\phi+\theta_1-\theta_3)}
\label{eq:X_mean_+0}
\end{equation}
and
\begin{equation}
\langle \hat{\sigma}_x \rangle_{-\alpha,0}=c_0c_2~\cos{(\phi+\theta_2-\theta_0)}+c_1c_3~\cos{(\phi+\theta_3-\theta_1)}
\label{eq:X_mean_-0}
\end{equation}
respectively.

(IV) For $\beta_{i}=0$, $\beta_{j}=\alpha_{i}$ and $\beta_{i}=0$, $\beta_{j}=-\alpha_{i}$ we find two more equations,
\begin{equation}
\langle \hat{\sigma}_x \rangle_{0,\alpha}=c_0c_1~\cos{(\phi+\theta_0-\theta_1)}+c_2c_3~\cos{(\phi+\theta_2-\theta_3)}.
\label{eq:X_mean_0+}
\end{equation}
and
\begin{equation}
\langle \hat{\sigma}_x \rangle_{0,-\alpha}=c_0c_1~\cos{(\phi+\theta_1-\theta_0)}+c_2c_3~\cos{(\phi+\theta_3-\theta_2)}
\label{eq:X_mean_0-}
\end{equation}

Finally for $\beta_\text{i,j}=0$ we obtain the following relation
\begin{equation}
\langle \hat{\sigma}_x \rangle_{0,0}=\left(c_0^2+c_1^2+c_2^2+c_3^2\right)~\cos{\phi},
\label{eq:X_mean_00}
\end{equation}
which is equivalent to the normalisation condition for $\ket{\Psi}_{\text{ij}}$ with the additional degree of freedom $\phi$.

The above equations are not yet in a form where they can be used to obtain all pairs of $c_i, \theta_i$ straightforwardly.
However, they can be combined and further simplified using basic trigonometric relations as shown below:

(i) First, adding and subtracting equations (\ref{eq:X_mean_++}) and (\ref{eq:X_mean_--}) we obtain
\begin{equation}
\langle \hat{\sigma}_x \rangle_{\alpha,\alpha}+\langle \hat{\sigma}_x \rangle_{-\alpha,-\alpha}=2c_0c_3~\cos{\phi}~\cos{(\theta_3-\theta_0)},
\end{equation}
and 
\begin{equation}
\langle \hat{\sigma}_x \rangle_{\alpha,\alpha}-\langle \hat{\sigma}_x \rangle_{-\alpha,-\alpha}=2c_0c_3~\sin{\phi}~\sin{(\theta_3-\theta_0)}.
\end{equation}
If the qubit is prepared such that $\phi=\pi/4$ then by combining the above two equations we obtain a relation for $c_0,~c_3$ that does not depend on $\theta_0,~\theta_3$:
\begin{equation}
c_0c_3=\sqrt{|\langle \hat{\sigma}_x \rangle_{\alpha,\alpha}|^2 +|\langle\hat{\sigma}_x \rangle_{-\alpha,-\alpha}|^2}.
\label{eq:c0c3}
\end{equation}
If $c_0c_3\neq0$ we can also determine the phases.
First, $e^{i\theta_0}$ in Eq.~(\ref{eq:Psi_general_SI}) can be absorbed into a global phase factor multiplying $\ket{\Psi}_{\text{ij}}$ followed by a redefinition of $\theta_{1,2,3}\rightarrow\theta_{1,2,3}/\theta_0$ (equivalent to defining $\theta_0=0$ or $2\pi$).
Then for $\phi=\pi/4$ we have
\begin{equation}
\theta_3=\arctan{\left(\frac{\langle \hat{\sigma}_x \rangle_{\alpha,\alpha}-\langle \hat{\sigma}_x \rangle_{-\alpha,-\alpha}}{\langle \hat{\sigma}_x \rangle_{\alpha,\alpha}+\langle \hat{\sigma}_x \rangle_{-\alpha,-\alpha}}\right)}.
\label{eq:theta_3}
\end{equation}
% \begin{equation}
% \frac{\langle \hat{\sigma}_x \rangle_{\alpha,\alpha}-\langle \hat{\sigma}_x \rangle_{-\alpha,-\alpha}}{\langle \hat{\sigma}_x \rangle_{\alpha,\alpha}+\langle \hat{\sigma}_x \rangle_{-\alpha,-\alpha}}=\tan{\phi}~\tan{\theta_3-\theta_0},
% \end{equation}

(ii) Following the same recipe we can obtain similar relations for $c_1,~c_2$ and $\theta_1,~\theta_2$.
In this case, by combining equations (\ref{eq:X_mean_+-}) and (\ref{eq:X_mean_-+}) for $\phi=\pi/4$ we obtain the following equations
\begin{equation}
c_1c_2=\sqrt{|\langle \hat{\sigma}_x \rangle_{\alpha,-\alpha}|^2 +|\langle\hat{\sigma}_x \rangle_{-\alpha,\alpha}|^2},
\label{eq:c1c2}
\end{equation}
and (assuming $c_1c_2\neq0$)
\begin{equation}
\theta_2-\theta_1=\arctan{\left(\frac{\langle \hat{\sigma}_x \rangle_{\alpha,-\alpha}-\langle \hat{\sigma}_x \rangle_{-\alpha,\alpha}}{\langle \hat{\sigma}_x \rangle_{\alpha,-\alpha}+\langle \hat{\sigma}_x \rangle_{-\alpha,\alpha}}\right)}.
\label{eq:theta_21}
\end{equation}

(iii) Furthermore, from equations (\ref{eq:X_mean_+0}) and (\ref{eq:X_mean_-0}) we obtain (for $\phi=\pi/4$)
\begin{align}
&\left(\langle \hat{\sigma}_x \rangle_{\alpha,0}+\langle \hat{\sigma}_x \rangle_{-\alpha,0}\right)^2\pm\left(\langle \hat{\sigma}_x \rangle_{\alpha,0}-\langle \hat{\sigma}_x \rangle_{-\alpha,0}\right)^2\nonumber \\
&~=2\left[(c_0c_2)^2+(c_1c_3)^2+2c_0c_1c_2c_3~\cos{(\theta_2\pm\theta_1\mp\theta_3)}\right].
\end{align}
Using equations (\ref{eq:c0c3}), (\ref{eq:theta_3}), (\ref{eq:c1c2}) and (\ref{eq:theta_21}) we can obtain a relation for $c_0,c_1,c_2,c_3$ with no dependence on the phases:
\begin{widetext}
\begin{align}
(c_0c_2)^2+(c_1c_3)^2&=f\left(\langle \hat{\sigma}_x \rangle_{\alpha,0}, \langle \hat{\sigma}_x \rangle_{-\alpha,0}, \langle \hat{\sigma}_x \rangle_{\alpha,\alpha}, \langle \hat{\sigma}_x \rangle_{-\alpha,-\alpha}, \langle \hat{\sigma}_x \rangle_{\alpha,-\alpha}, \langle \hat{\sigma}_x \rangle_{-\alpha,\alpha}\right)\nonumber\\
&=2\langle \hat{\sigma}_x \rangle_{\alpha,0}\langle \hat{\sigma}_x \rangle_{-\alpha,0} - 2\Biggl[\sqrt{\left(|\langle \hat{\sigma}_x \rangle_{\alpha,\alpha}|^2 +|\langle\hat{\sigma}_x \rangle_{-\alpha,-\alpha}|^2\right) \left(|\langle \hat{\sigma}_x \rangle_{\alpha,-\alpha}|^2 +|\langle\hat{\sigma}_x \rangle_{-\alpha,\alpha}|^2\right)} \nonumber\\
&~~~~~~~~~~~~\times\cos{\left(\arctan{\left(\frac{\langle \hat{\sigma}_x \rangle_{\alpha,\alpha}-\langle \hat{\sigma}_x \rangle_{-\alpha,-\alpha}}{\langle \hat{\sigma}_x \rangle_{\alpha,\alpha}+\langle \hat{\sigma}_x \rangle_{-\alpha,-\alpha}}\right)}+\arctan{\left(\frac{\langle \hat{\sigma}_x \rangle_{\alpha,-\alpha}-\langle \hat{\sigma}_x \rangle_{-\alpha,\alpha}}{\langle \hat{\sigma}_x \rangle_{\alpha,-\alpha}+\langle \hat{\sigma}_x \rangle_{-\alpha,\alpha}}\right)}\right)}\Biggr].
\label{eq:c0c2_c1c3}
\end{align}
\end{widetext}

(iv) Similarly, from equations (\ref{eq:X_mean_0+}) and (\ref{eq:X_mean_0-}) we obtain (for $\phi=\pi/4$)
\begin{align}
&\left(\langle \hat{\sigma}_x \rangle_{0,\alpha}+\langle \hat{\sigma}_x \rangle_{0,-\alpha}\right)^2\pm\left(\langle \hat{\sigma}_x \rangle_{0,\alpha}-\langle \hat{\sigma}_x \rangle_{0,-\alpha}\right)^2\nonumber\\
&~=2\left[(c_0c_1)^2+(c_2c_3)^2+2c_0c_1c_2c_3~\cos{(\theta_1\pm\theta_2\mp\theta_3)}\right].
\end{align}
Again, using equations (\ref{eq:c0c3}), (\ref{eq:theta_3}), (\ref{eq:c1c2}) and (\ref{eq:theta_21}) we can obtain another relation for $c_0,c_1,c_2,c_3$ with no dependence on the phases:
\begin{widetext}
\begin{align}
(c_0c_1)^2+(c_2c_3)^2&=g\left(\langle \hat{\sigma}_x \rangle_{0,\alpha}, \langle \hat{\sigma}_x \rangle_{0,-\alpha}, \langle \hat{\sigma}_x \rangle_{\alpha,\alpha}, \langle \hat{\sigma}_x \rangle_{-\alpha,-\alpha}, \langle \hat{\sigma}_x \rangle_{\alpha,-\alpha}, \langle \hat{\sigma}_x \rangle_{-\alpha,\alpha}\right)\nonumber\\
&=2\langle \hat{\sigma}_x \rangle_{0,\alpha}\langle \hat{\sigma}_x \rangle_{0,-\alpha} - 2\Biggl[\sqrt{\left(|\langle \hat{\sigma}_x \rangle_{\alpha,\alpha}|^2 +|\langle\hat{\sigma}_x \rangle_{-\alpha,-\alpha}|^2\right) \left(|\langle \hat{\sigma}_x \rangle_{\alpha,-\alpha}|^2 +|\langle\hat{\sigma}_x \rangle_{-\alpha,\alpha}|^2\right)} \nonumber\\
&~~~~~~~~~~~~\times\cos{\left(\arctan{\left(\frac{\langle \hat{\sigma}_x \rangle_{\alpha,\alpha}-\langle \hat{\sigma}_x \rangle_{-\alpha,-\alpha}}{\langle \hat{\sigma}_x \rangle_{\alpha,\alpha}+\langle \hat{\sigma}_x \rangle_{-\alpha,-\alpha}}\right)}-\arctan{\left(\frac{\langle \hat{\sigma}_x \rangle_{\alpha,-\alpha}-\langle \hat{\sigma}_x \rangle_{-\alpha,\alpha}}{\langle \hat{\sigma}_x \rangle_{\alpha,-\alpha}+\langle \hat{\sigma}_x \rangle_{-\alpha,\alpha}}\right)}\right)}\Biggr].
\label{eq:c0c1_c2c3}
\end{align}
\end{widetext}

In our case we are interested in reading out the Bell state
\begin{equation}
\ket{\Psi}_\text{ij}=\frac{1}{\sqrt{N}}\left(\ket{0_{i} 0_{j}} + e^{i\theta} \ket{\alpha_{i}\alpha_{j}}\right),
\label{Psi_target_SI}
\end{equation}
i.e. the state in Eq.~(\ref{eq:Psi_general_SI}) with $\theta_3=\theta$, $c_0=c_3=\frac{1}{\sqrt{N}}$ and $c_1=c_2=0$.
Let us assume that we have prepared the general state in Eq.~(\ref{eq:Psi_general_SI}).
First, we can measure $\langle \hat{\sigma}_x \rangle_{\alpha,\alpha}$ and $\langle \hat{\sigma}_x \rangle_{-\alpha,-\alpha}$ and from Eq.~(\ref{eq:c0c3}) determine $c_0c_3$.
If we have indeed prepared the target state shown in Eq.~(\ref{Psi_target_SI}) then this product should be nonzero.
Then we proceed by measuring $\langle \hat{\sigma}_x \rangle_{\alpha,-\alpha}$ and $\langle \hat{\sigma}_x \rangle_{-\alpha,\alpha}$ which should both be zero indicating that either $c_1=0$ or $c_2=0$ according to Eq.~(\ref{eq:c1c2}).
Additionally equations~(\ref{eq:c0c2_c1c3}) and~(\ref{eq:c0c1_c2c3}) should also equate to zero indicating $c_1=c_2=0$.
Finally, combining equations~(\ref{eq:X_mean_00}) and~(\ref{eq:c0c3}) we have
\begin{equation}
(c_0-c_3)^2=\sqrt{2}\langle \hat{\sigma}_x \rangle_{0,0}-2\sqrt{|\langle \hat{\sigma}_x \rangle_{\alpha,\alpha}|^2 +|\langle\hat{\sigma}_x \rangle_{-\alpha,-\alpha}|^2}.
\label{eq:ReadoutBell_SI}
\end{equation}
If indeed the state in Eq.~(\ref{Psi_target_SI}) is prepared then we should find that $\langle \hat{\sigma}_x \rangle_{0,0}=\sqrt{2\left(|\langle \hat{\sigma}_x \rangle_{\alpha,\alpha}|^2 +|\langle\hat{\sigma}_x \rangle_{-\alpha,-\alpha}|^2\right)}$, and therefore $c_0=c_3=\left(|\langle \hat{\sigma}_x \rangle_{\alpha,\alpha}|^2 +|\langle\hat{\sigma}_x \rangle_{-\alpha,-\alpha}|^2\right)^{1/4}=2^{-1/4}\langle \hat{\sigma}_x \rangle_{0,0}$.

% {\color{red}We can calculate it numerically and compare to the expected value of $1/2\sqrt{2}$. $\langle \hat{\sigma}_x \rangle_{0,0}$ should be equal to $1/2^6$}

% Solving the system of equations (\ref{eq:X_mean_00}), (), (\ref{eq:c0c2_c1c3}) and (\ref{eq:c0c1_c2c3})
% \sqrt{|\langle \hat{x} \rangle_{\alpha,-\alpha}|^2 +|\langle\hat{x} \rangle_{-\alpha,\alpha}|^2}

\bibliography{../BibMarios}

\end{document}